\documentclass[a4paper,fleqn]{cas-dc}
\usepackage{booktabs}
\usepackage{amsmath}
\usepackage{wrapfig}
\usepackage{subfig}
 \usepackage{subcaption}
\usepackage[accsupp]{axessibility}  
\usepackage{bm}
\usepackage{multirow}
\usepackage{tikz}
\usepackage{comment}
\usepackage{amsmath,amssymb} 
\usepackage{color}
\usepackage{pifont}
\usepackage{makecell}
\usepackage{caption}
\usepackage{float}
\usepackage{adjustbox}
\usepackage{etoolbox}
\usepackage{dsfont}

\usepackage[accsupp]{axessibility}  
\usepackage{paralist}

\def\eg{\textit{e.g.}} 
\def\ie{\textit{i.e}.} 

\newcommand{\PAR}[1]{\noindent{\bf #1}}

\usepackage{amsmath,lipsum}
\usepackage{cuted}
\usepackage{lineno}
\usepackage{booktabs,makecell,multirow}
\usepackage{amsmath,amssymb,amsfonts}
\usepackage{algorithmic}
\usepackage{graphicx}
\usepackage{textcomp}
\usepackage{amsmath,amsfonts}
\usepackage{algorithmic}
\usepackage{algorithm}
\usepackage{array}
\usepackage{textcomp}
\usepackage{stfloats}
\usepackage{url}
\usepackage{verbatim}
\usepackage{graphicx}
\usepackage{cite}
\usepackage{booktabs}
\usepackage{paralist}

\usepackage{threeparttable}

\usepackage[numbers,sort&compress]{natbib}

\def\tsc#1{\csdef{#1}{\textsc{\lowercase{#1}}\xspace}}
\tsc{WGM}
\tsc{QE}

\usepackage{xspace}
\makeatletter
\DeclareRobustCommand\onedot{\futurelet\@let@token\@onedot}
\def\@onedot{\ifx\@let@token.\else.\null\fi\xspace}
\def\eg{\textit{e.g}\onedot} 
\def\ie{\textit{i.e}\onedot}

\makeatother

\usepackage[table]{xcolor} 
\definecolor{Gray}{gray}{0.9} 
\definecolor{rblue}{rgb}{0,0.5,1}

\newcommand{\JQ}[1]{\textcolor{black}{#1}}

\usepackage{xcolor}
\definecolor{hollywoodcerise}{rgb}{0.96, 0.0, 0.63}
\definecolor{lasallegreen}{rgb}{0.03, 0.47, 0.19}
\definecolor{hanpurple}{rgb}{0.32, 0.09, 0.98}
\definecolor{green(pigment)}{rgb}{0.0, 0.65, 0.31}

\hypersetup{colorlinks=true,linkcolor={red},citecolor={hanpurple},urlcolor={magenta}} 

\usepackage{caption}
\begin{document}
\let\WriteBookmarks\relax

\shorttitle{OmniLens: Towards Universal Lens Aberration Correction via LensLib-to-Specific Domain Adaptation}    

\shortauthors{Q. Jiang \textit{et al.}}  

\title [mode = title]{OmniLens: Towards Universal Lens Aberration Correction via LensLib-to-Specific Domain Adaptation}

\author[1]{Qi Jiang}
\cormark[1]
\credit{Conceptualization of this study, Methodology, Software}
\author[1]{Yao Gao}
\cormark[1]
\author[1]{Shaohua Gao}
\author[1]{Zhonghua Yi}
\author[1]{Xiaolong Qian}
\author[1]{Hao Shi}
\author[2,3]{Kailun Yang} 
\author[1,4]{Lei Sun}[orcid=0000-0001-7310-5565]
\cormark[2]
\ead{leo_sun@zju.edu.cn}
\author[1]{Kaiwei Wang}[orcid=0000-0002-8272-3119]
\cormark[2]
\ead{wangkaiwei@zju.edu.cn}
\author[1]{Jian Bai}

\affiliation[a]{organization={State Key Laboratory of Extreme Photonics and Instrumentation, College of Optical Science and Engineering},
            addressline={Zhejiang University}, 
            city={Hangzhou},
            postcode={310027}, 
            country={China}}

\affiliation[b]{organization={School of Artificial Intelligence and Robotics},
            addressline={Hunan University}, 
            city={Changsha},
            postcode={410012}, 
            country={China}}

\affiliation[c]{organization={National Engineering Research Center of Robot Visual Perception and Control Technology},
addressline={Hunan University}, 
city={Changsha},
postcode={410082}, 
country={China}}

\affiliation[d]{organization={Sofia University St. ``Kliment Ohridski''},
            addressline={INSAIT}, 
            city={Sofia},
            postcode={1784}, 
            country={Bulgaria}}

%

\cortext[1]{Equal contribution}
\cortext[2]{Corresponding author}

\begin{abstract} 
Emerging universal Computational Aberration Correction (CAC) paradigms provide an inspiring solution to light-weight and high-quality imaging with a universal model trained on a lens library (LensLib) to address arbitrary lens optical aberrations blindly. However, the limited coverage of existing LensLibs leads to poor generalization of the trained models to unseen lenses, whose fine-tuning pipeline is also confined to the lens-descriptions-known case. In this work, we introduce \textit{OmniLens}, a flexible solution to universal CAC via (i) establishing a convincing LensLib with comprehensive coverage for pre-training a robust base model, and (ii) adapting the model to any specific lens designs with unknown lens descriptions via fast LensLib-to-specific domain adaptation.
To achieve these, an Evolution-based Automatic Optical Design (EAOD) pipeline is proposed to generate a rich variety of lens samples with realistic aberration behaviors. Then, we design an unsupervised regularization term for efficient domain adaptation on a few easily accessible real-captured images based on the statistical observation of dark channel priors in degradation induced by lens aberrations. Extensive experiments demonstrate that the LensLib generated by EAOD effectively develops a universal CAC model with strong generalization capabilities, which can also improve the non-blind lens-specific methods by 0.35${\sim}$1.81dB in PSNR. Additionally, the proposed domain adaptation method significantly improves the base model, especially in severe aberration cases (at most 2.59dB in PSNR). The code and data will be available at \url{https://github.com/zju-jiangqi/OmniLens}.
\end{abstract}

\maketitle

\section{Introduction}
\label{sec:intro}
\begin{figure}
  \centering
  \includegraphics[width=1.0\linewidth]{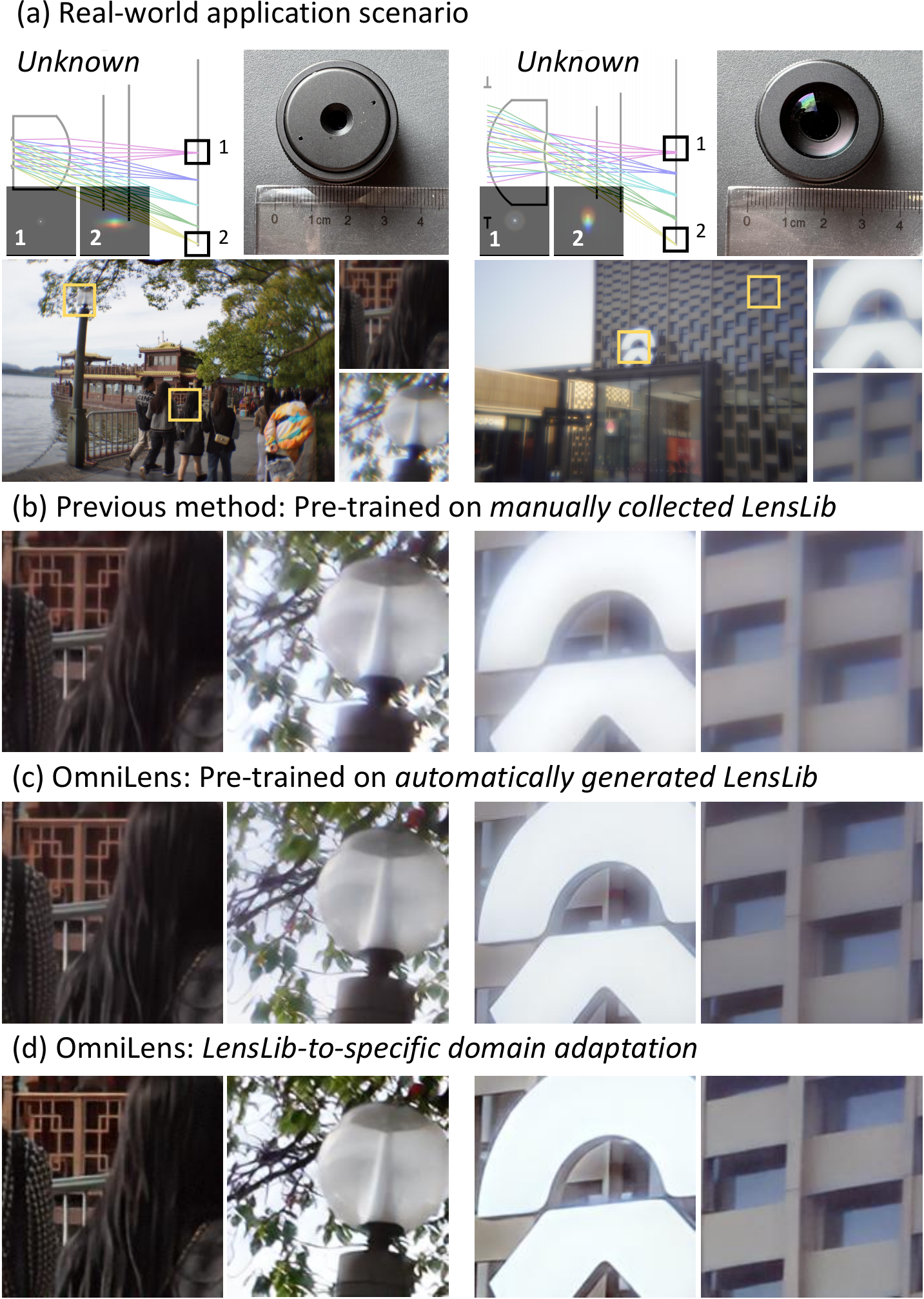}
  \caption{\textbf{OmniLens is a flexible solution to universal CAC.} In real-world cases of unknown lens descriptions for the applied low-end lenses, the pre-trained model with manually collected LensLib~\cite{li2021universal} delivers poor performance and can hardly be further fine-tuned. In contrast, the automatically generated LensLib in OmniLens develops a more robust model, where the LensLib-to-specific domain adaptation can further enhance the results.}
  \label{fig:teaser}
\end{figure}

Computational Aberration Correction (CAC), equipped with a post-image-restoration method to deal with the degradation (\ie, the optical degradation~\cite{chen2021optical}) induced by residual optical aberrations of the target optical lens, is a fundamental and long-standing task~\cite{schuler2011non,kee2011modeling} in computational imaging.
This technology is particularly highly demanded in mobile and wearable vision terminals for lightweight and high-quality photography~\cite{suo2023computational,zhang2023large}, where low-end lenses~\cite{heide2013high,gao2022compact} with simple structure and severe optical degradation are applied, as shown in Fig.~\ref{fig:teaser} (a).  
Recent advances in CAC have centered around Deep Learning (DL) based  methods~\cite{lin2022non,chen2021extreme_quality,chen_mobile_2023} equipped with a powerful image restoration network~\cite{liang2021swinir, chen2022simple, zamir2022restormer} being trained on the corresponding data pairs under lens-specific optical degradation~\cite{chen2021extreme_quality,chen2021optical}. 
However, these methods have invariably been tailored to a specific lens, revealing limited generalization ability to other different lens designs.
In this way, the complex and time-consuming pipeline of lens description calibration, imaging simulation, data preparation, and model training requires re-conduction for every new lens, limiting the applications of lens-specific CAC methods.

Consequently, developing a universal CAC model trained on datasets under a lens library (LensLib) to address arbitrary optical degradation has hit the forefront.
Yet, existing solutions~\cite{li2021universal, gong2024physics} suffer two main limitations:
(i) \textit{the coverage of the manually collected LensLib is limited and inflexible for expanding}, leading to poor zero-shot CAC results when the target lens appears as a new, previously unseen sample, as shown in Fig.~\ref{fig:teaser} (b);
(ii) \textit{the fine-tuning strategy~\cite{li2021universal} for model adaptation requires known or calibrated lens descriptions for data preparation}, which can hardly be achieved in real-world applications.
Recently, the blooming development of Domain Adaptation (DA)~\cite{wang2021unsupervised,xu2022dual,wang2023domain} provides an effective pipeline to adapt the model trained on sufficient labeled data to unlabeled data with domain shift, which reveals potential in tuning the universal CAC model for a description-unknown lens, but has not been explored.

In this work, we build up a flexible universal CAC framework OmniLens, modeling the CAC as the problem of domain adaptation from a large LensLib to specific lenses.
First and foremost, a convincing LensLib covering possible aberration behaviors of real-world lenses is first constructed to pre-train a base model with strong generalization capabilities.
To achieve this, instead of manual design or collection, we propose a novel Evolution-based Automatic Optical Design (EAOD) method for automatic LensLib generation, consisting of an evolution mechanism for enriching the diversity of the generated lens structures, and comprehensive optimization objectives to ensure their reasonable aberration behaviors.
As shown in Fig.~\ref{fig:teaser} (c), the model pre-trained on automatically generated LensLib generalizes well to the training-unseen lenses.
Then, incorporating the easily accessible real-captured images of a specific lens, the LensLib-to-specific domain adaptation is conducted to fine-tune the pre-trained model for better specific CAC results. 
To this intent, we develop an efficient DA framework based on our statistical observation of Dark Channel Prior (DCP)~\cite{pan2016blind} in optical degradation, which is expressed as an unsupervised regularization term to constrain the CAC model for fast and few-shot domain adaptation.
Despite being in a blind deconvolution way without access to the lens descriptions, the adapted model can achieve excellent results as shown in Fig.~\ref{fig:teaser} (d).

Extensive experiments under various testing lenses and networks further demonstrate that:
i) The LensLib constructed by EAOD delivers state-of-the-art performance in training a robust blind universal CAC model;
ii) The LensLib-based pre-training further improves the non-blind lens-specific model;
iii) The LensLib-to-specific domain adaptation brings significant improvements to the base model, especially in tough and real-world cases.

In summary, OmniLens offers a solution for users without an optical background to CAC of an unknown lens, which also provides a robust pre-training foundation for research on non-blind lens-specific CAC. 
The main contributions are summarized as follows:
\begin{compactitem}
    \item We deliver a novel insight into universal CAC, modeling it as LensLib-to-specific domain adaptation. 
    \item We design the Evolution-based Automatic Optical Design (EAOD) method to build up a convincing LensLib, developing a pre-trained foundation with strong generalization ability. 
    \item An efficient and unsupervised DA framework is proposed to adapt the pre-trained model to specific lenses.
    \item Extensive experiments across $6$ testing lenses and $3$ networks verify the effectiveness of OmniLens for correcting diverse lens aberrations blindly with unknown lens descriptions, where the pre-trained foundation can also facilitate non-blind lens-specific solutions.
\end{compactitem}

\section{Related Work}
\label{sec:related}
\begin{figure*}[!t]
  \centering
  \includegraphics[width=1.0\linewidth]{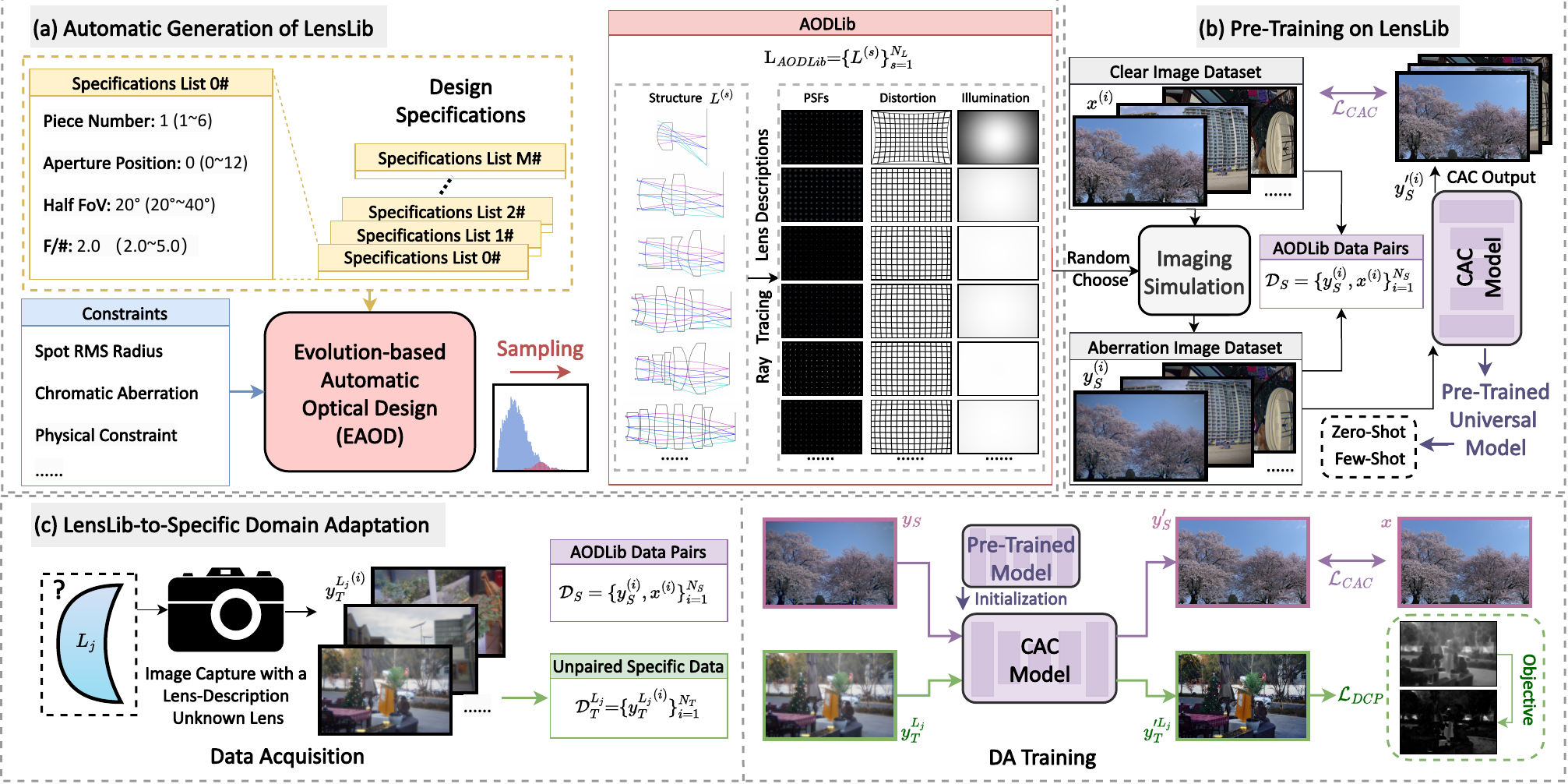}
  \caption{\textbf{Overview of the established OmniLens framework.} (a): The generation of AODLib by the proposed EAOD method with multiple design specifications and constraints fed in. (b) Based on the AODLib data pairs prepared by imaging simulation, we pre-train a universal CAC model for zero-shot or few-shot CAC of target lenses. (c): Incorporating the real-world captured images of the target lens-description unknown lens, LensLib-to-specific domain adaptation is conducted to fine-tune the pre-trained model for enhancing the results, where the DA training is guided by an unsupervised loss derived from DCP of the optical degradation. }
  \label{fig:overview}
\end{figure*}

\noindent\textbf{Computational Aberration Correction.}
Computational Aberration Correction (CAC)~\cite{schuler2011non,heide2013high} is proposed to enhance the images of low-end lenses with optical degradation.
Benefiting from the blooming development of image restoration networks in low-level vision~\cite{liang2021swinir,chen2022simple,chen2022real}, learning-based CAC methods~\cite{peng2019learned,chen2021extreme_quality,chen_mobile_2023} have revealed more impressive results than model-based methods~\cite{Schuler2012Blind,yue2015blind,eboli2022fast}. 
These methods can be regarded as an implicit modeling of end-to-end non-blind deconvolution, wherein the model learns the deconvolution for a specific lens using data generated from calibrated or simulated lens descriptions.
Yet, limited to lens-specific solutions, these models can hardly accommodate new unseen lenses. 
Recently, universal CAC~\cite{li2021universal,gong2024physics} has been explored for addressing the optical degradation of any lens blindly, where the generalization ability of the model highly depends on the coverage of the LensLib used for training.
Existing LensLib construction centered around manually collected lens samples~\cite{li2021universal,gong2024physics}, \ie, ZEBASELib, or random Zernike model~\cite{hu2021image,jiang2024computational}, \ie, ZernikeLib.
However, the former suffers from the limited number and coverage, and the latter overlooks the domain shift between generated virtual lenses and real-world ones.
Moreover, their fine-tuning pipelines are often not applicable in lens-descriptions-unknown scenarios. 
Consequently, we propose OmniLens to address these limitations, constructing a more convincing LensLib by automatic optical design, and a LensLib-to-specific domain adaptation pipeline to fine-tune the model with unknown lens descriptions.

\noindent\textbf{Automatic Optical Design.}
Heuristic algorithms, \eg, Simulated Annealing (SA), Genetic Algorithm (GA), and Particle Swarm Optimization (PSO) have gained widespread adoption in the field of automatic optical design (AOD)~\cite{guo2019new, sun2021automatic, zhang2020automated, yue2022adaptive, tang2019ant, reichert2020development} for searching the optimized lens structure under the given constraints. 
Recently, Yang~\textit{et al.} have also proposed to combine curriculum learning and deep optics for achieving better AOD results~\cite{yang2023curriculum}.
However, existing works only focus on finding a certain optimal optical system, ignoring the exploration of possible diverse structures.
Efforts have been made to pursue diverse design solutions using Deep Neural Networks (DNN)~\cite{cote2019extrapolating, cote2021deep}. 
Yet, their performance is restricted to the training lens database.
Moreover, the lens structures output by the network are not regularized by any stringent quality or physical constraints, which can hardly cover the lenses in practical applications.
\JQ{Heuristic algorithms that consider image quality and physical constraints have recently been used to search diverse initial structures in end-to-end computational imaging systems~\cite{gao2025exploring}. 
The genetic algorithm-based evolution mechanism plays a significant role, which drives overall optimization toward a global optimum while also searching a large number of intermediate local optima. 
Motivated by this, considering that these intermediate local optima are precisely aberration cases that may occur in real-world lens optimization, we put forward the EAOD method based on heuristic algorithms to generate a diverse range of lenses automatically with machinable structures and reasonable aberration behaviors to cover real-world lenses.}

\noindent\textbf{Domain Adaptation in Low-Level Vision.}
Early effort has been made in mitigating the synthetic-to-real gap problem in CAC via domain adaptation from a labeled synthetic domain to the unlabeled target real-world domain~\cite{jiang2025representing}, which demands a large amount of data and training iterations for learning the domain-mixing representation of optical degradation.
Nevertheless, the optical degradation of LensLib varies across each training sample, which can hardly be represented in a uniform way mixed with the target-specific domain. 
The heavy training cost also renders such approaches unsuitable for the required rapid model fine-tuning in our framework.
In addition, there have been works on domain adaptation in other low-level vision tasks, \eg, image super-resolution~\cite{wei2021unsupervised,wang2021unsupervised}, image dehazing~\cite{li2019semi,chen2021psd,li2023dadrnet}, and underwater image enhancement~\cite{jiang2022two,wang2023domain}. 
Among them, we find that the utilization of task-specific physical priors (\eg, DCP)~\cite{chen2021psd,li2019semi} contributed to effective adaptation with minimal training cost.
To achieve this, we verify the \JQ{existence} of DCP in optical degradation, deriving it into a regularization term to enable the LensLib-to-specific domain adaptation.

\section{Methodology}
\label{sec:method}
\begin{figure*}[!t]
  \centering
  \includegraphics[width=1.0\linewidth]{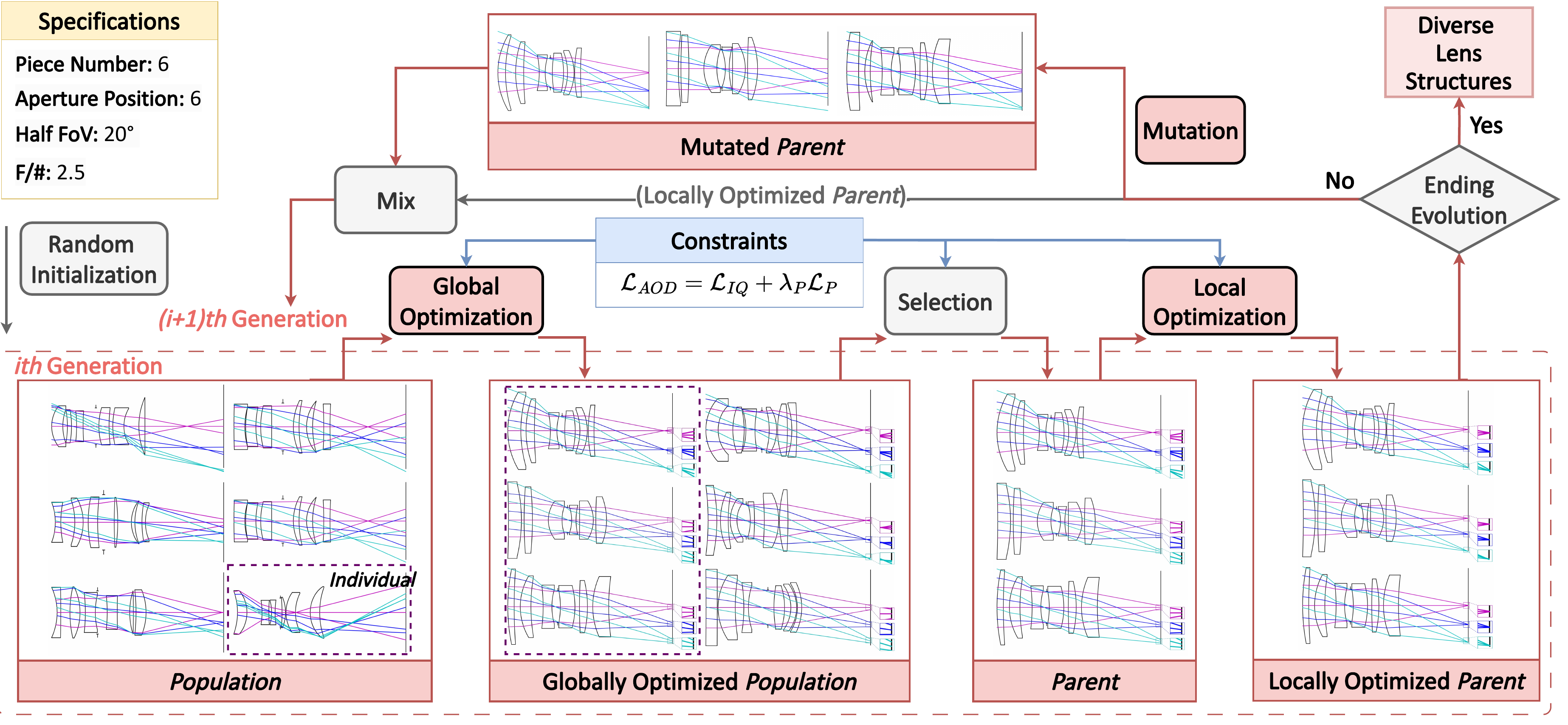}
  \caption{\textbf{Overall pipeline of the proposed EAOD method.} Taking the lens specifications of $6$ piece as an example, the EAOD leverages a hybrid global and local optimization strategy to seek diverse lens structures that maximally satisfy both imaging quality and physical constraints from generation to generation based on an evolution framework with a mutation mechanism.}
  \label{fig:EAOD}
\end{figure*}

\subsection{\JQ{Motivation}}
\JQ{With recent advances in large datasets and models, image restoration has begun to move toward an all‑in‑one paradigm~\cite{jiang2025survey,potlapalli2023promptir,ai2024multimodal}.} 
\JQ{These developments indicate that training on broadly distributed multiple image degradations enables a universal model to effectively learn many‑to‑one mappings from degradations such as noise, motion blur, rain, fog, and low light to clear images. In addition, in common learning‑based non‑blind CAC pipelines, models are typically trained on aberration data with multiple combinations of perturbations~\cite{chen2022computational_mass,jiang2024minimalist} to enhance generalization to unknown manufacturing errors. This strategy effectively prototypes the universal paradigm and shows that training over multiple distributions enables handling diverse aberrations. 
In this way, this paper explores to construct a large data source covering possible real-world aberration behaviors and pre-train a universal model for aberration correction. 
Meanwhile, given that generalization is always bounded, we also investigate efficient fine-tuning pipelines based on the pre-trained universal model as a remedy for zero‑shot failure cases, forming a comprehensive and flexible universal aberration correction framework.}

\subsection{Framework Overview}
\label{sec:overview}
A flexible two-stage framework OmniLens is introduced for universal CAC in real-world applications with unknown lens descriptions.
As shown in Fig.~\ref{fig:overview} (a)$\sim$(b), the first stage is the \textit{pre-training of a universal model on simulated data pairs under a LensLib}, where the core part is to generate possibly diverse and realistic lens samples by the EAOD method for constructing a convincing LensLib.
To deal with the possible failure cases of the pre-trained model in a zero-shot manner confronted with unseen lenses, we propose a novel \textit{LensLib-to-specific domain adaptation} pipeline for the CAC of the target-specific lens at the second stage. 
As shown in Fig.~\ref{fig:overview} (c), the additional data required is solely a small number of real-world captured images by the target lens, which is easily accessible and requires no lens descriptions.
The DA training is conducted based on the AODLib data pairs and the unpaired specific data, which is guided by both the unsupervised regularization derived from the observed DCP in optical degradation and the supervised loss.

\subsection{Automatic Generation of LensLib}
\label{sec:lib}
Considering the highly free and empirical process of optical design, designing a large number of lenses manually to traverse possible aberrations of real-world lenses is challenging. 
To tackle this issue, the idea of AOD~\cite{cote2021deep} is introduced to enable the large-scale generation of lens samples.

\noindent\textbf{Modeling the Automatic Optical Design.}
Overall, AOD aims to seek optimized lens structures under one given specifications set $\mathcal{S}^{(j)}$.
To broaden the coverage of the designed lenses, four types of specifications that deliver large impacts on the final aberration behavior, \ie, the piece number $S_{p}$, aperture position $S_{aper}$, half FoV $S_{FoV}$ and F number $S_{F}$ of the lens, are considered for composing $\mathcal{S}^{(j)}$.
Then, we characterize a lens structure by its curvatures, surface parameters, glass and air spacings, refractive indexes, and Abbe numbers, defining a normalized lens parameters vector, whose dimension and value range meet the given specifications $\mathcal{S}^{(j)}{=}\{S_{p}^{(j)}, S_{aper}^{(j)}, S_{FoV}^{(j)}, S_{F}^{(j)}\}$: 
\begin{equation}
\label{eq:lensvector}
P^{\mathcal{S}^{(j)}}=(p^{(1)}, p^{(2)},...,p^{(n)})^T \in \mathbb{R}^n.
\end{equation}
The optimization process of $P^{\mathcal{S}^{(j)}}$ is guided by an objective function $\mathcal{L}_{AOD}(P^{\mathcal{S}^{(j)}})$, consisting of both imaging quality constraints and physical structure constraints.
Specifically, an imaging quality loss $\mathcal{L}_{IQ}$ calculated by ray-tracing~\cite{chen2021optical} is applied as the main optimization objective, which can be the common RMS radius or Modulation Transfer Function (MTF) loss in traditional optical design\cite{steinich2012optical}.
Moreover, key physical properties of the lens derived from $P^{\mathcal{S}^{(j)}}$ are constrained by a linear penalty to fall within a reasonable range: 
\begin{equation}
\label{eq:lp}
\mathcal{L}_{P} = \dfrac{1}{n_{P}}\sum^{n_{P}}_{i=1}\alpha_i[\max(q^{(i)}_{min} - q_i, 0)+\max(q_i - q^{(i)}_{max} , 0)],
\end{equation}
where $n_{P}$ numbers of physical quantities $q_i$ are constrained between $q^{(i)}_{min}$ and $q^{(i)}_{max}$, and the constraint ratio is regulated by weight $\alpha_i$.
We employ $\mathcal{L}_{P}$ to ensure machinable output structures while meeting the specifications, in particular preventing adjacent surfaces from being too close together or even overlapped, which is a common error in the optical design process. Finally, $\mathcal{L}_{AOD}$ considering both constraints with a balancing weight $\lambda_{P}$ is expressed as:
\begin{equation}
\label{eq:lEAOD}
\mathcal{L}_{AOD} = \mathcal{L}_{IQ} + \lambda_{P}\mathcal{L}_{P}.
\end{equation}
Please refer to the supplemental document for more details about the implementation of $\mathcal{L}_{AOD}$.
Finally, the AOD process can be modeled as:
\begin{equation}
\label{eq:AODfinal}
\hat{{P}}^{\mathcal{S}^{(j)}}=\arg\min\mathcal{L}_{AOD}(P^{\mathcal{S}^{(j)}}),
\end{equation}
where $\hat{{P}}^{\mathcal{S}^{(j)}}$ is the optimized lens structure. 

\noindent\textbf{Evolution-based Automatic Optical Design.}
Based on the Eq.~(\ref{eq:AODfinal}), we propose an EAOD method to (i) generate more possible lens structures under one set of specifications, and (ii) ensure the best imaging quality for the current structure for driving the aberration behaviors of the generated samples closer to those of real-world ones. 

As shown in Fig.~\ref{fig:EAOD}, EAOD builds upon a Genetic Algorithm (GA)~\cite{katoch2021review} based evolution framework to produce optimized lens structures from generation to generation.
We first define a $Population$ composed of all lens structures to be optimized, where each structure is coined an $Individual$. 
Without any empirical preference, the $Individual$ is the normalized lens parameters vector $P^{\mathcal{S}^{(j)}}$ from random initialization, and $m$ $Individuals$ are initialized to constitute the $Population$ for enriching diversity:
\begin{equation}
\label{eq:init}
\mathrm{P}^{\mathcal{S}^{(j)}}= \{ P^{\mathcal{S}^{(j)}}_1, P^{\mathcal{S}^{(j)}}_2 ,..., P^{\mathcal{S}^{(j)}}_{m} \}.
\end{equation}
Then, a hybrid global and local optimization strategy is proposed to find multiple promising high-quality structures $\hat{\mathrm{P}}^{\mathcal{S}^{(j)}}$ for $\mathrm{P}^{\mathcal{S}^{(j)}}$, coined the $Parent$, which minimizes the $\mathcal{L}_{AOD}$.
Concretely, the global optimization method, \eg, SA~\cite{rutenbar1989simulated} or PSO~\cite{985692}, is employed to rapidly drive the randomly initialized $Individuals$ towards the location of the global optimal solution, where the solutions with superior $\mathcal{L}_{AOD}$ are selected for the following local optimization; while the local optimization method, \eg, ADAM~\cite{kingma2014adam} or Damped Least Squares (DLS)~\cite{levenberg1944method}, aims to seek their local optima for the relatively best imaging results under the primarily optimized structures.
In addition, $Parent$ is mutated for more possible structures, where for each $Individual$ ${\hat{{P}}^{\mathcal{S}^{(j)}}}_{i}$ in $Parent$, $\gamma$ percentage of its parameters are randomly initialized.
The mutated $Parent$ is then mixed with the origin $Parent$ to serve as the start points of the next generation.
Finally, the hybrid optimization, selection, and mutation processes are re-conducted for each generation, where the $Parent$ is outputted as the diverse lens structures when the generation meets the set number.

\noindent\textbf{AODLib Construction.}
\label{sec:sampling}
With all sets of specifications $\mathrm{S}{=}\{\mathcal{S}^{(j)}\}_{j=1}^{M}$ fed into EAOD, the initial LensLib $\mathrm{L}_{lib}$ is constructed.
We further quantize its aberration behavior by using the average RMS spot radii (${\mu}{m}$) of a lens sample ($RMS$ for short), which can intuitively reflect the severity of the samples' aberrations.
To mitigate the data preference towards any certain level of aberration, we propose to sample a portion of lenses from the $\mathrm{L}_{lib}$ to construct the final AODLib ${\mathrm{L}}_{AODLib}$:
\begin{equation}
\label{eq:sampling}
\mathrm{L}_{AODLib}{=}\{L^{(s)}\}_{s=1}^{N_S} = Sampling(\mathrm{L}_{lib}, r_{min}, r_{max}, N_S),
\end{equation}
where we hope to obtain $N_{S}$ samples with a uniform distribution of the $RMS$ within the range $[r_{min}, r_{max}]$.

\begin{figure}
  \centering
  \includegraphics[width=0.8\linewidth]{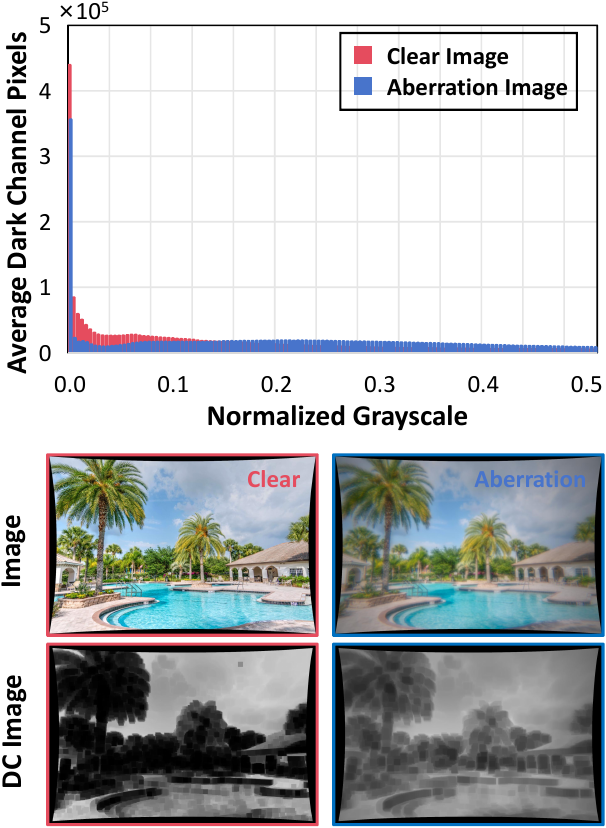}
  \caption{\textbf{Illustration of dark channel prior in CAC.} (a): The grayscale histogram of the DC images for aberration-clear image pairs. (b): The corresponding visual samples.}
  \label{fig:dcp}
\end{figure}

\subsection{Pre-Training on LensLib}
We employ a ray-tracing model~\cite{chen2021optical} to calculate the lens descriptions for all samples in ${\mathrm{L}}_{AODLib}$, which are applied to simulate aberration images via patch-wise convolution with clear images and distortion transformation. 
In this way, the AODLib data pairs $\mathcal{D}_{S}{=}\{{y_S}^{(i)}, {x}^{(i)}\}_{i=1}^{N}$ are prepared for supervised pre-training using the objective $\mathcal{L}_{CAC}({y'_S}^{(i)}, {x}^{(i)})$, where ${y'_S}^{(i)}$ is the CAC output. 
$\mathcal{L}_{CAC}$ can be any training loss in the CAC field, such as the combination of L2 loss and perceptual loss~\cite{chen2021extreme_quality}, depending on the applied CAC model.
The pre-trained universal model can be applied to correct aberrations of real-world lenses in a zero-shot manner.
In addition, when the target lens description is available, it can serve as the pre-trained foundation for improving the lens-specific CAC model during few-shot fine-tuning or full model training.

\subsection{LensLib-to-Specific Domain Adaptation}
\label{sec:uda}
In this stage, few aberration images $\mathcal{D}^{L_{j}}_{T}{=}\{{y^{L_{j}}_{T}}^{(i)}\}_{i=1}^{N_T}$ captured by the target lens $L_{j}$ are utilized to adapt the pre-trained model trained on $D_S$ (the source domain) to $\mathcal{D}^{L_{j}}_{T}$'s optical degradation domain (the target domain).

Effective low-level vision DA frameworks~\cite{shao2020domain,chen2021psd,jiang2022two} are often built upon the statistical priors of degraded images or clean images, \eg, the Dark Channel Prior (DCP)~\cite{he2010single} in image dehazing, which can provide an unsupervised regularization for model training.
Motivated by the observation from~\cite{pan2016blind} that convolution-induced blur often exhibits DCP, we hypothesize that optical degradation also adheres to DCP, as it originates from the convolution with the PSFs.
To validate the hypothesis, we use the DIV2K dataset~\cite{timofte2017ntire} of natural images, and without loss of generality, choose the second lens in Fig.~\ref{fig:teaser} to construct aberration-clear image pairs. 
The Dark Channel (DC) image $D(I)$ for an image $I$ can be calculated via a minimum value filtering operation: 
\begin{equation}
\label{eq:dci}
D(I) = \min\limits_{y\in\mathcal{N}(x)}(\min\limits_{c\in\{r,g,b\}}I^c(y)),
\end{equation}
where $I^c$ is the $c$th color channel, $x,y$ are pixel locations, and $\mathcal{N}(x)$ represents an image patch centered at $x$. 
We statistically analyze the intensity distribution histograms of the $D(I)$ for all the aforementioned aberration-clear image pairs as shown in Fig.~\ref{fig:dcp}, where a set of visual samples of $I$ and $D(I)$ is also provided. 
Corroborating our hypothesis, the elements in DC images for clear natural images mostly tend towards zero, while aberration images exhibit far more non-zero pixels in their DC images, indicating that optical degradation indeed possesses DCP (further experimental evidences are provided in Sec.~\ref{exp:dcp}).

Based on the observation, we derive the DCP into $\mathcal{L}_{DCP}$ as an unsupervised constraint on the CAC output ${y'^{L_{j}}_{T}}^{(i)}$ of the target lens $L_{j}$:
\begin{equation}
\label{eq:dcploss}
\mathcal{L}_{DCP} = \Vert{D({y'^{L_{j}}_{T}}^{(i)})}\Vert_1,
\end{equation}
where the $D({y'^{L_{j}}_{T}}^{(i)})$ will be pulled towards zero to align with the dark channel distribution of natural clear images.
To improve the stability of the DA training, we also adopt $\mathcal{L}_{CAC}$ to provide supervision on the source domain AODLib data $\mathcal{D}_S$, constructing the overall training objective $\mathcal{L}_{DA}$:
\begin{equation}
\label{eq:udaloss}
\mathcal{L}_{DA} = \lambda_{S}\mathcal{L}_{CAC} + \lambda_{T}\mathcal{L}_{DCP},
\end{equation}
where $\lambda_{S}$ and $\lambda_{T}$ are loss weights.

\section{Experiments and Results}
\label{sec:exp}

\begin{table}[t!]
    \begin{center}
        \caption{\textbf{Quantitative comparison with different LensLibs.} 
        For conciseness, we only list the average performance of all test lenses. The detailed results can be found in the supplemental document. We highlight the \textbf{best} and \underline{second best} results.}
        \label{tab:lib}
        \resizebox{0.5\textwidth}{!}
{

\renewcommand{\arraystretch}{1.55}
\setlength{\tabcolsep}{2mm}{
\begin{tabular}{cccccccc}
\bottomrule[0.17em]
\multirow{2}{*}{\textbf{LensLib}} & \multirow{2}{*}{\textbf{Network}} & \multirow{2}{*}{\textbf{Params}} & \multicolumn{5}{c}{\textbf{Average Performance}} \\ \cline{4-8} 
 &  &  & \textbf{PSNR$\uparrow$} & \textbf{SSIM$\uparrow$} & \textbf{LPIPS$\downarrow$} & \textbf{FID$\downarrow$}& \textbf{OS$\uparrow$} \\ \hline \hline
\multirow{3}{*}{\begin{tabular}[c]{@{}c@{}}ZEBASELib\\\cite{li2021universal,gong2024physics}\end{tabular}} & FOV-KPN &4.01M  &24.01  &0.748  &0.2547  &41.22 &63.13  \\
 & SwinIR &11.97M  &23.54  &0.755  &0.2618  &41.07 &62.82 \\
 & FeMaSR &37.37M  &23.55  &0.737  &0.2288  &38.76 &64.28 \\ \hline
\multirow{3}{*}{\begin{tabular}[c]{@{}c@{}}ZernikeLib\\\cite{jiang2024computational,hu2021image}\end{tabular}} & FOV-KPN &4.01M  &25.16  &0.776  &0.1910  &33.56 &68.76\\
 & SwinIR &11.97M  &24.45  &0.784  &0.1788  &29.20 &69.98 \\
 & FeMaSR &37.37M  &23.73  &0.758  &0.1620  &30.98  &69.49\\ \hline \hline
\multirow{3}{*}{\begin{tabular}[c]{@{}c@{}}AODLib\\-LensNet~\cite{cote2021deep}\end{tabular}} & FOV-KPN &4.01M  &23.84  &0.750  &0.2466  &41.09 &63.54 \\
 & SwinIR &11.97M  &24.79  &0.773  &0.2271  &36.94 &66.21 \\
 & FeMaSR &37.37M  &23.79  &0.748  &0.2097  &36.41 &66.01 \\ \hline
\multirow{3}{*}{\textbf{\begin{tabular}[c]{@{}c@{}}AODLib\\-EAOD (Ours)\end{tabular}}} & FOV-KPN &4.01M  &\underline{25.62}  &0.785  &0.1902  &27.16 &70.21 \\
 & SwinIR &11.97M  &\textbf{26.13}  &\textbf{0.817}  &\underline{0.1622}  &\textbf{20.82} &\textbf{73.73} \\
 & FeMaSR &37.37M  &25.13  &\underline{0.788}  &\textbf{0.1409}  &\underline{22.88} &\underline{73.22} \\ \hline
\end{tabular}

}
}
    \end{center}
\end{table}

\begin{figure}
  \centering
  \includegraphics[width=1.0\linewidth]{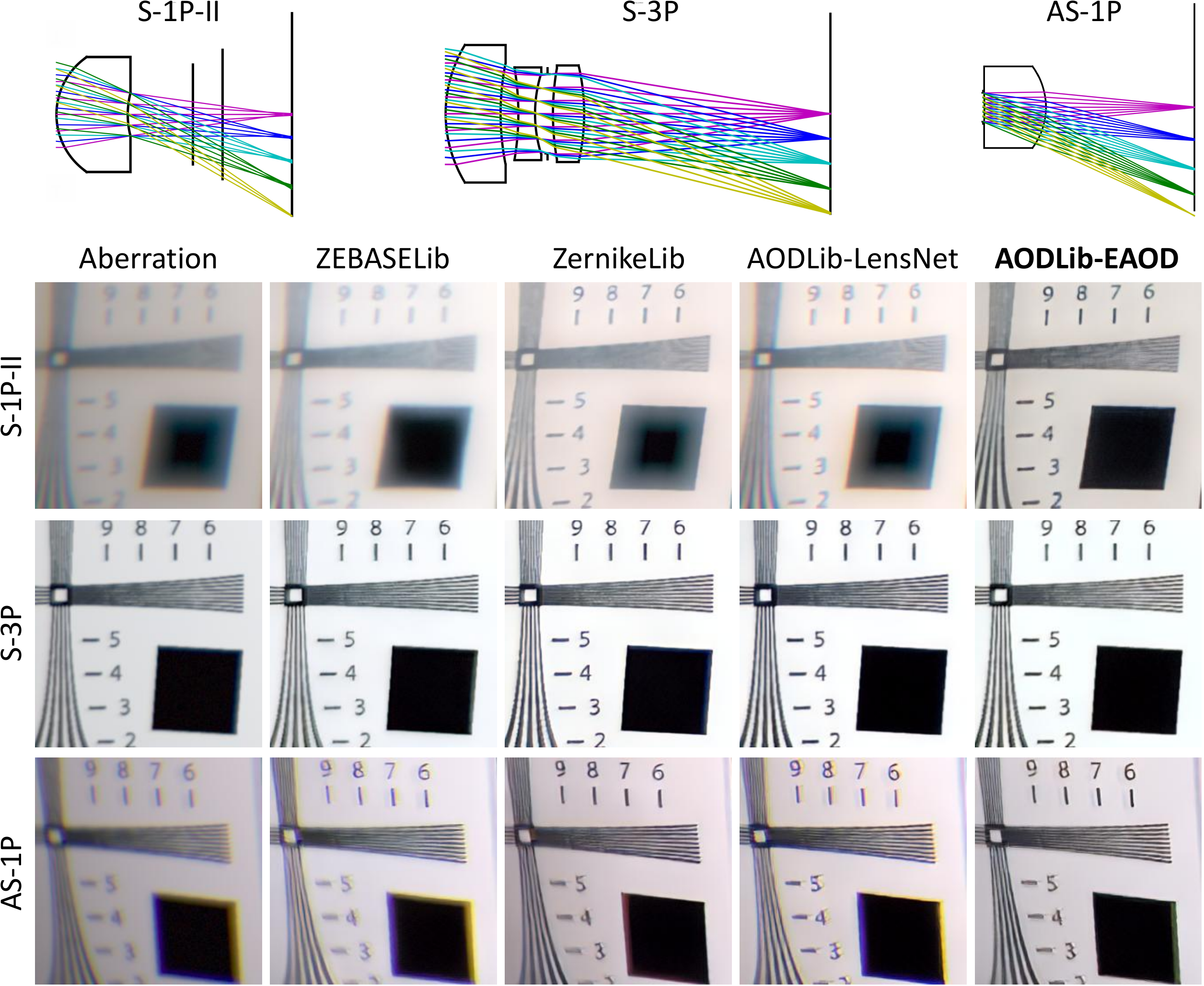}
  \caption{\textbf{Visual comparison with different LensLibs under $3$ representative test lenses, which cover different surface types and levels of aberrations.} The optical paths of the $3$ lenses are shown at the top of the figure. For each LensLib solution, we present its visual CAC results on an ISO12233 chart produced by FeMaSR which delivers the best perceptual performance.}
  \vskip-4ex
  \label{fig:visual_lib}
\end{figure}

\begin{table}[t!]
    \begin{center}
        \caption{\textbf{Quantitative evaluation of the OmniLens framework.} 
        ``*'' denotes the additional data and iterations for training the lens-specific model. Similarly read as Tab.~\ref{tab:lib}.}
        \label{tab:main}
        \resizebox{0.5\textwidth}{!}
{
\renewcommand{\arraystretch}{1.25}
\setlength{\tabcolsep}{1mm}{
\begin{tabular}{cccccccc}
\bottomrule[0.17em]
\multirow{2}{*}{\textbf{Solution}} & \multirow{2}{*}{\textbf{Network}} & \multirow{2}{*}{\textbf{\begin{tabular}[c]{@{}c@{}}Training Cost*\\ (Imgs/Iters)\end{tabular}}} & \multicolumn{5}{c}{\textbf{Average Performance}} \\ \cline{4-8} 
 &  &  & \textbf{PSNR$\uparrow$} & \textbf{SSIM$\uparrow$} & \textbf{LPIPS$\downarrow$} & \textbf{FID$\downarrow$}& \textbf{OS$\uparrow$} \\ \hline \hline
\multicolumn{7}{c}{\textit{Lens-Descriptions-Unknown (Blind Deconvolution)}}\\ \hline
 Eboli \textit{et al.} & - &0/0K  &22.07  &0.689  &0.3562  &49.55&54.26  \\ \hline
\multirow{3}{*}{\textbf{\begin{tabular}[c]{@{}c@{}}OmniLens\\ Zero-Shot\end{tabular}}} & FOV-KPN & \multirow{3}{*}{0/0K} &25.62  &0.785  &0.1902  &27.16&70.21\\
 & SwinIR &  &\underline{26.13}  &\textbf{0.817}  &0.1622  &\underline{20.82}&73.73  \\
 & FeMaSR &  &25.13  &0.788  &0.1409  &22.88&73.22  \\ \hline
\multirow{3}{*}{\textbf{\begin{tabular}[c]{@{}c@{}}OmniLens\\DA\end{tabular}}}  & FOV-KPN   & \multirow{3}{*}{40/10K} &25.76  &0.783  &0.1845  &24.84&70.84 \\
 & SwinIR &    &\textbf{26.18}  &\underline{0.816}  &0.1538  &\textbf{20.03}&\underline{74.26}  \\
 & FeMaSR &    &25.61  &0.792  &\textbf{0.1260}  &\textbf{20.03}&\textbf{74.70}  \\ \hline\hline

\multicolumn{7}{c}{\textit{Lens-Descriptions-Known (Non-Blind Deconvolution)}}\\ \hline
\multicolumn{7}{c}{\textit{Few-Shot}}\\ \hline
\multirow{3}{*}{\begin{tabular}[c]{@{}c@{}}Specific\end{tabular}} & FOV-KPN & \multirow{3}{*}{105/10K} & 23.97  &0.764  &0.2247  &28.62&66.98  \\
 & SwinIR &   &25.35  &0.800  &0.1911  &22.04&71.28  \\
 & FeMaSR &   &24.11  &0.768  &0.1499  &23.85&71.61  \\ \hline
\multirow{3}{*}{\textbf{\begin{tabular}[c]{@{}c@{}}Specific\\+OmniLens\end{tabular}}} & FOV-KPN & \multirow{3}{*}{105/10K} &\underline{26.05}  &\underline{0.804}  &0.1613  &18.95&73.63  \\
 & SwinIR &  &\textbf{26.86}  &\textbf{0.836}  &\underline{0.1386}  &\textbf{15.68}&\underline{76.54}  \\
 & FeMaSR &  &25.96  &0.801  &\textbf{0.1077}  &\underline{15.84} &\textbf{76.65} \\ \hline\hline
\multicolumn{7}{c}{\textit{Full Training}}\\ \hline
\multirow{3}{*}{\begin{tabular}[c]{@{}c@{}}Specific\end{tabular}} & FOV-KPN & \multirow{3}{*}{2111/200K} &26.42  &0.810  &0.1541  &18.14&74.44  \\
 & SwinIR &   &\underline{26.77}  &\underline{0.837}  &0.1436  &15.99&76.24  \\
 & FeMaSR &   &25.54  &0.804  &\underline{0.1149}  &17.00 &76.04 \\ \hline 
\multirow{3}{*}{\textbf{\begin{tabular}[c]{@{}c@{}}Specific\\ +OmniLens\end{tabular}}} & FOV-KPN & \multirow{3}{*}{2111/200K} &26.61  &0.817  &0.1496  &17.42&75.06  \\
 & SwinIR &  &\textbf{27.09}  &\textbf{0.841}  &0.1356  &\underline{15.18}&\underline{77.01}  \\
 & FeMaSR &  &26.08  &0.807  &\textbf{0.1010}  &\textbf{14.59}&\textbf{77.40}  \\ \hline
\end{tabular}

}
}
    \end{center}
\end{table}

\begin{figure}
  \centering
  \includegraphics[width=1.0\linewidth]{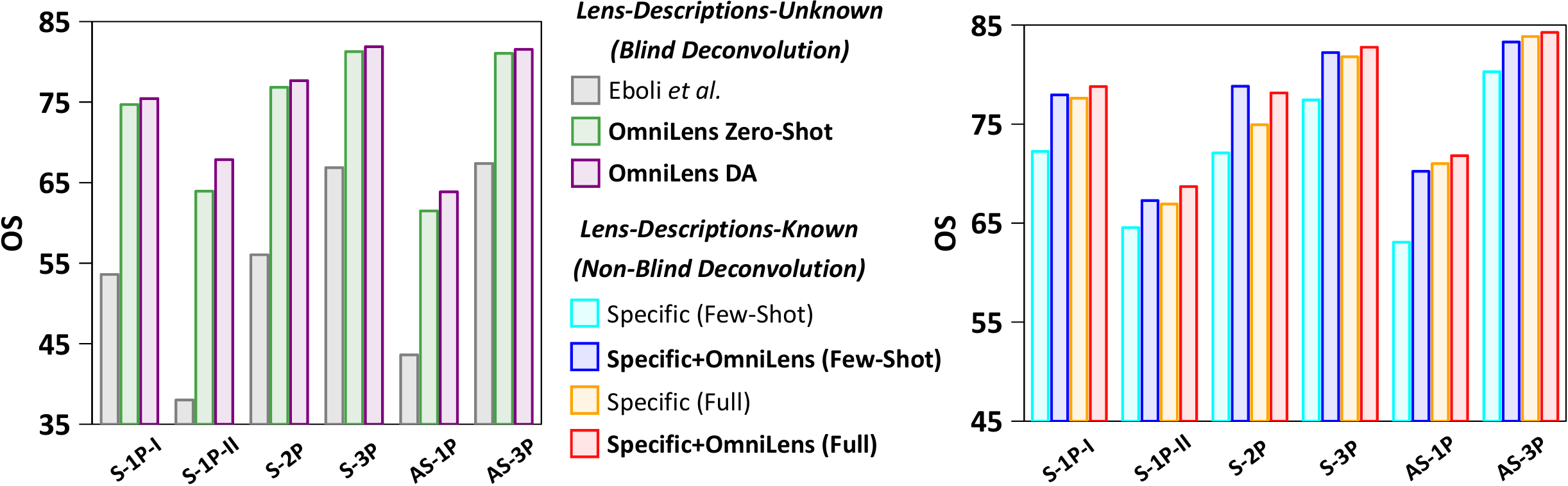}
  \caption{\textbf{Evaluation of OmniLens across all test lenses.} We show the OS results produced by FeMaSR with the best performance.}
  \label{fig:ridar}
\end{figure}

\begin{figure}
  \centering
  \includegraphics[width=1.0\linewidth]{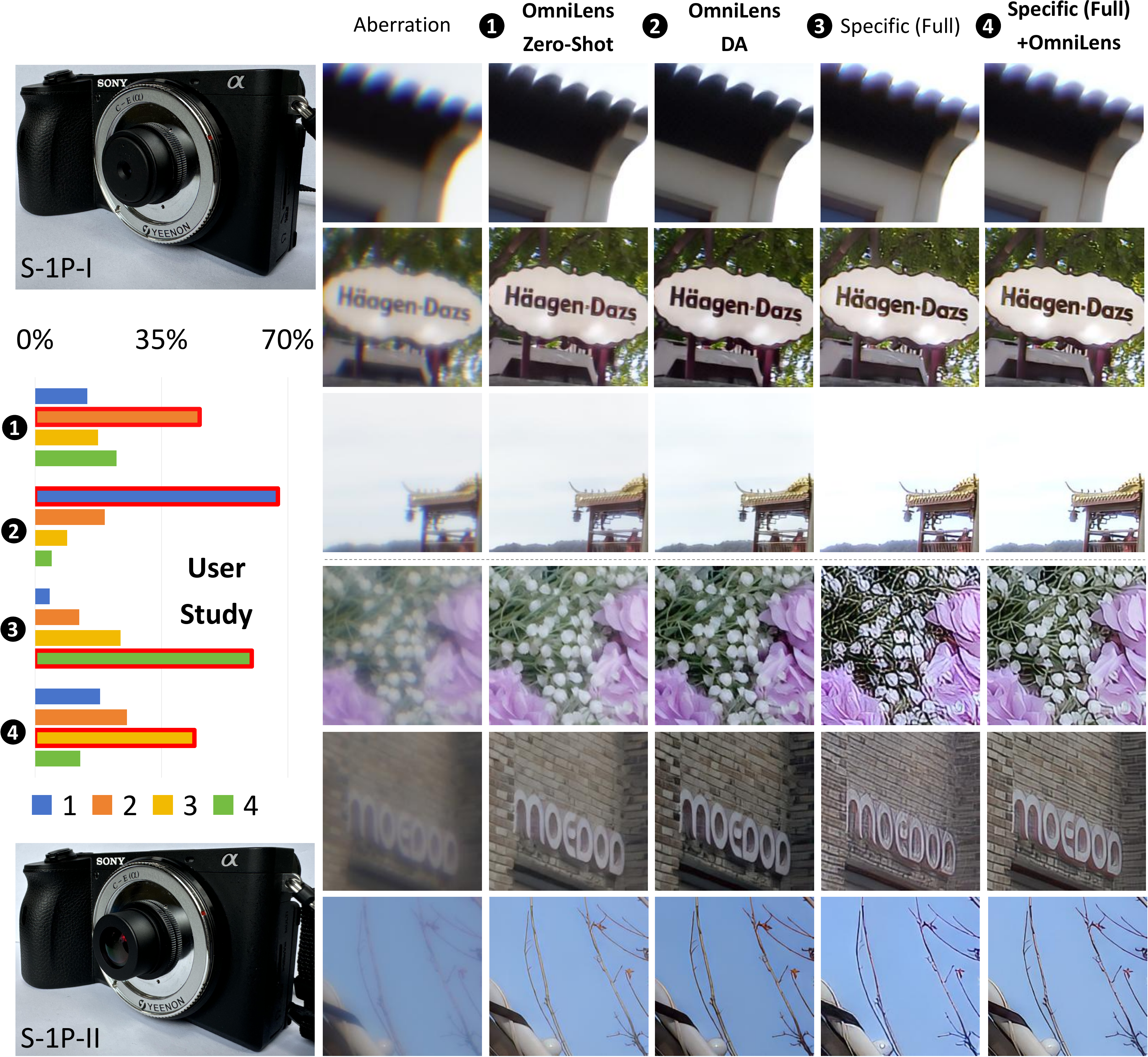}
  \caption{\textbf{Results on the real-world benchmark.} The shooting device for capturing real-world images is shown on the left of the figure, where the fabricated S-1P-I and S-1P-II are applied. We also present the results of a user study (detailed in the supplement), showing the percentage of each method ranked from first to fourth.}
  \label{fig:visual_real}
\end{figure}

\subsection{Implementation Details}

\noindent\textbf{Evaluation Protocol.}
We manually design $6$ different specific lenses as the test samples, including $4$ spherical lenses (S-1P-I, S-1P-II, S-2P, and S-3P) and $2$ aspherical lenses (AS-1P and AS-3P) of different piece numbers. 
A synthetic benchmark of simulated images under each lens is then set up based on imaging simulation, where the referenced metrics, \ie, PSNR, SSIM~\cite{wang2004image}, LPIPS~\cite{zhang2018unreasonable}, and FID~\cite{heusel2017gans} are employed for comprehensive evaluation. 
To intuitively evaluate the performance of each solution, we normalize and weight these metrics following~\cite{liang2024ntire}, constructing an Overall Score (OS):
\begin{equation}
\label{eq:os}
\begin{split}
&OS = 20\times\frac{PSNR}{50} + 15\times\frac{SSIM-0.5}{0.5}\\ 
& + 20\times\frac{1-LPIPS}{0.4}+ 15\times\frac{100-FID}{100}.
\end{split}
\end{equation}
In addition, we fabricate two of the single spherical lenses to capture test images for constructing a real-world benchmark.
A commercial lens: $CAYE_\circledR 50mm$, under different F-numbers, is also applied for further evaluation.

\noindent\textbf{Data Preparation.}
The Flickr2K ~\cite{timofte2017ntire} is chosen to generate $D_{S}$.
In terms of synthetic evaluation, to avoid information leakage, target-specific training data $\mathcal{D}^{L_{j}}_{T}$ and test data are prepared with another dataset DIV2K to simulate the real-world captured images, where the ground-truth is unavailable during DA training.
For real-world evaluation, we capture $50$ images as the test set, and another $25$ images as $\mathcal{D}^{L_{j}}_{T}$, with each aforementioned fabricated lens. 

\noindent\textbf{AODLib Construction.}
Considering the optimization complexity, we only generate the spherical lens samples to ensure the quality of the optimized structures.
In terms of EAOD, to balance optimization capability and convergence speed, we apply SAA and ADAM for the global optimization and local optimization respectively.

\noindent\textbf{Model Training.}
We select FOV-KPN~\cite{chen2021extreme_quality} (CNN and spatially-varying design), SwinIR~\cite{liang2021swinir} (Transformer), and FeMaSR~\cite{chen2022real} (CNN, Transformer, and generative model) as the representative networks in CAC to evaluate our OmniLens framework, all trained with default learning rates and loss functions for $200K$ iterations on $D_{S}$ to obtain the pre-training model.
While for DA training, we use $L_{DA}$ with a window size of $15$ for DC image calculation to fine-tune the pre-trained model for $10K$ iterations with a learning rate of $1e{-}6$, where $\lambda_{S}$ and $\lambda_{T}$ are set to $0.01$ and $1$ respectively, which is consistent across all $3$ networks.

More details can be found in the supplemental document.

\subsection{Comparison with Different LensLibs}
As shown in Tab.~\ref{tab:lib}, we first verify the effectiveness of AODLib built by EAOD via comparison with the other two LensLib pipelines, \ie, the ZEBASELib and the ZernikeLib introduced in Sec.~\ref{sec:related}, whose construction process is consistent with~\cite{gong2024physics} and~\cite{jiang2024computational} respectively.
Additionally, the deep-learning-based AOD method LensNet~\cite{cote2021deep} is also applied to construct a competing AODLib.
The universal models with different architectures trained on our AODLib-EAOD outperform those under other LensLibs by a large margin, \eg, achieving the best OS, and bringing average gains of $1.78dB{\sim}1.93dB$ in PSNR and ${-}0.084{\sim}{-}0.013$ in LPIPS.
It can also be seen from the visual results in Fig.~\ref{fig:visual_lib} that the CAC results of other pipelines still suffer from optical degradation under severe aberrations, where those of our method are sharp and clear.
These evidences prove that the proposed EAOD method can build a more convincing LensLib for developing a robust universal CAC model generalizing well to most real-world lenses.

\subsection{Evaluation of OmniLens}
In lens-descriptions-unknown cases, we compare our pre-trained model and DA model in OmniLens with the state-of-the-art blind deconvolution method Eboli \textit{et al.}~\cite{eboli2022fast}.
For cases with known lens descriptions, lens-specific models under the selected $3$ networks are selected as baselines to evaluate  OmniLens pre-training.

\noindent\textbf{Synthetic Benchmark.}
Tab.~\ref{tab:main} shows the average performance of competing methods in the synthetic benchmark, where Fig.~\ref{fig:ridar} also provides detailed overall results across each test lens. 
In the lens-descriptions-unknown situation, the proposed OmniLens significantly outperforms the traditional blind deconvolution method. Moreover, DA effectively improves the pre-trained model by $0.22dB$ in PSNR and $-0.0096$ in LPIPS.
The improvements are observed across most test cases, which are more obvious under lenses with severe aberrations, \eg, S-1P-I, S-1P-II, and AS-1P.
Then, with known lens descriptions, OmniLens pre-training brings improvements to the lens-specific models under both few-shot ($1.81dB$ in PSNR and $-0.0527$ in LPIPS) and full training ($0.35dB$ in PSNR and $-0.0088$ in LPIPS) settings.
Notably, even the few-shot specific model with OmniLens outperforms the full-trained one, illustrating OmniLens's potential to serve as a CAC pre-training foundation.

\begin{figure*}[h]
        \centering
        \includegraphics[width=\linewidth]{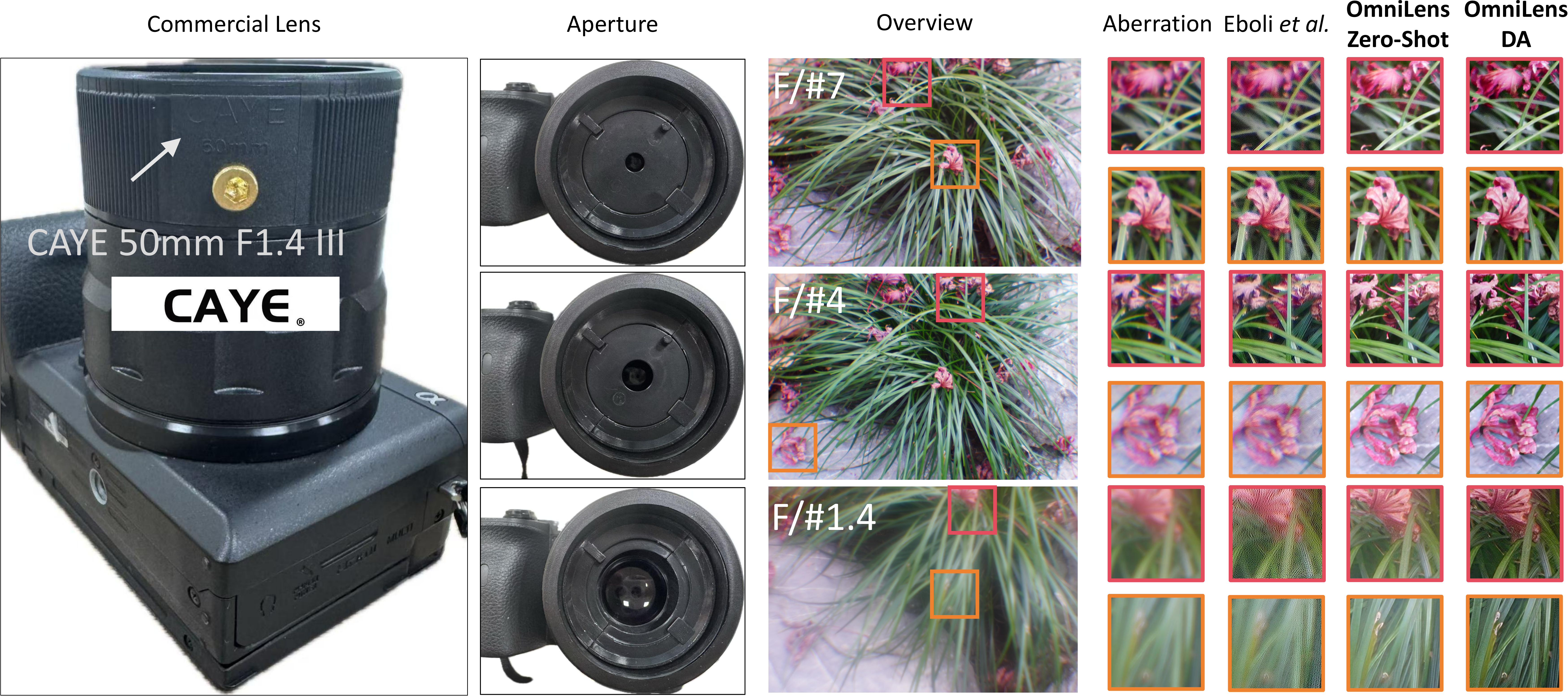} 
        \caption{\textbf{Application case of OmniLens on a commercial lens with unknown lens descriptions.} Results under different F-numbers with diverse aberration behaviors are provided for a comprehensive evaluation.}
        \label{fig:commercial}
\end{figure*}

\noindent\textbf{Real-world Benchmark.}
Fig.~\ref{fig:visual_real} presents some of the results (top $4$ solutions in Tab.~\ref{tab:main}) on the real-world benchmark.
Due to the synthetic-to-real domain gap, the results of the lens-specific solution (\textcircled{3}) reveal severe artifacts and unresolved optical degradation, while our OmniLens pre-training (\textcircled{1} and \textcircled{4}) can effectively suppress the artifacts. 
Moreover, OmniLens-DA (\textcircled{2}) shows great advantages in this case, effectively handling the optical degradation (image blur and purple fringes) without introducing visually unpleasant artifacts.
Benefiting from the regularization of $\mathcal{L}_{CAC}$ on AODLib, the DA method can work in sky regions where DCP fails.
The results of the user study are highly consistent with the visual results, where the results of OmniLens-based solutions are preferred by the majority of subjects.

\begin{figure*}[h]
    \centering
        \centering
        \includegraphics[width=\linewidth]{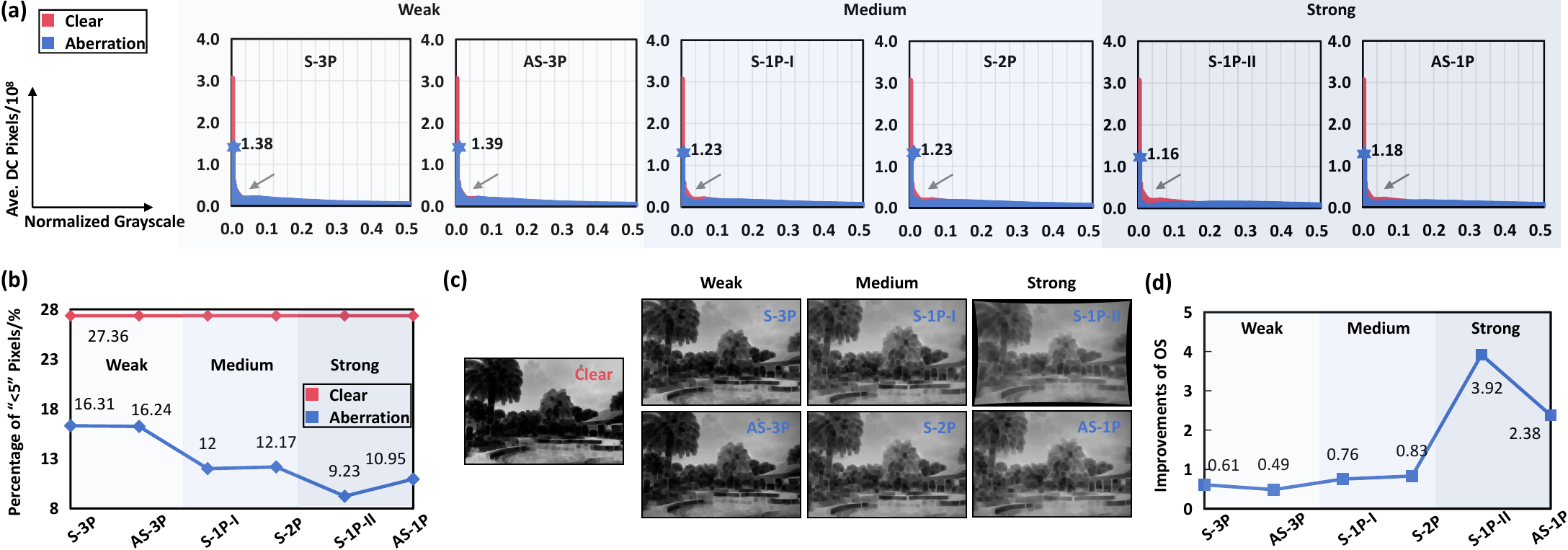} %
        \caption{\textbf{Detailed analysis of DCP in optical degradation.} (a): The grayscale histograms for the DC images of the $6$ test lenses. We mark the exact number of zero pixels corresponding to each lens aberration. (b): For each lens’s DC images, we report the proportion of pixels with grayscale values less than ``5'' relative to the total number of pixels. We also provide the same statistics for the clear images as a reference. (c): Visual results of DC images under different severities of aberrations. (d): OS improvements brought to the pre-trained model by the DCP-based DA under different lenses.}
        \label{fig:dcp_detail}
\end{figure*}

\noindent\textbf{Evaluation on the Commercial Lens.}
We further provide real-world results under a commercial lens: $CAYE_\circledR 50mm$ in Fig.~\ref{fig:commercial}, where different apertures are applied for diverse F-number settings.
Given that such a commercial lens lacks specific lens descriptions, we compare only the blind paradigm, namely the SOTA blind deconvolution method Eboli \textit{et al.}~\cite{eboli2022fast} and our OmniLens-Zero-shot and OmniLens-DA.
The results show the OmniLens pre-trained model effectively handles aberrations of such a commercial lens with an unknown lens description under large F-numbers.
Meanwhile, the DA method further enhances CAC results, especially in severe aberrations with a small F-number.
These cases with severe aberrations cannot be addressed by the traditional deconvolution method Eboli \textit{et al.}, indicating the favorable generalization of the OmniLens framework.

\subsection{Evaluation of DCP in Optical Degradation}
\label{exp:dcp}

To analyze the DCP properties of optical degradation under different severities of aberrations, the $6$ test lenses are first divided into ``weak'', ``medium'', and ``strong'' aberration levels (detailed in the supplemental document).
Using the same settings as the DCP analysis in Sec~\ref{sec:uda}, we conduct a statistical analysis of the DC images for all $6$ test lenses, as shown in Fig~\ref{fig:dcp_detail}.
It can be observed that aberrations at all levels exhibit DCP, which is reflected in dark channel images with fewer zero pixels (Fig.~\ref{fig:dcp_detail} (a)), a smaller proportion of pixels with small gray values (Fig.~\ref{fig:dcp_detail} (b)), and ``brighter'' DC images (Fig.~\ref{fig:dcp_detail} (c)), illustrating the university of DCP in optical degradation.
Meanwhile, the more severe aberrations reveal more obvious DCP properties. 
The experimental results in Fig.~\ref{fig:dcp_detail} (d) where more significant improvements in OS brought by DCP-based DA for ``strong'' aberrations further validate this.

\subsection{Ablation Study}
\label{exp:ab}
Without loss of generality, ablation studies are conducted with FeMaSR, presenting the average PSNR and LPIPS.

\begin{table}[h!]
    \begin{center}
        \caption{\textbf{Ablations on Sampling Strategy.} We mainly investigate the sampling distribution and number $N_S$ here, where more ablations can be found in the supplemental.}
        \label{tab:ab_sam}
        \resizebox{0.5\textwidth}{!}
{

\renewcommand{\arraystretch}{1.25}
\setlength{\tabcolsep}{4mm}{
\begin{tabular}{ccc|cc}
\bottomrule[0.17em]
& \textbf{Distribution} & \textbf{\bm{$N_{S}$}} & \textbf{PSNR$\uparrow$} & \textbf{LPIPS$\downarrow$} \\ \hline\hline
\textit{w./o. Sampling} & \textit{n.a.} & \textit{n.a.} &25.61  &0.1287  \\
1 & Gaussian & 100 &25.55  &0.1385  \\
2  & uniform & 100 &25.59  &0.1330  \\
3 & uniform & 10 &23.94  &0.1592  \\
4 & \textbf{uniform} & \textbf{1000} &\textbf{25.86}  &\textbf{0.1214}  \\ \hline
\end{tabular}

}
}

    \end{center}
\end{table}

\begin{table}[h!]
    \begin{center}
        \caption{\textbf{Ablations on EAOD.} The zero-shot results of the pre-trained model under each setting are reported. Specification settings: $S_p{=}6$ for \textit{w./o.} $S_p$, $S_{aper}{=}2$ for \textit{w./o.} $S_{aper}$, $S_{FoV}{=}20^\circ$ for \textit{w./o.} $S_{FoV}$, and $S_{F}{=}3.5$ for \textit{w./o.} $S_{F}$. Phy. Con.: Physical Constraints. Whether to employ local optimization is not ablated as it is indispensable for optimizing reasonable lens structures.}
        \label{tab:ab_aod}
        \resizebox{0.5\textwidth}{!}
{

\renewcommand{\arraystretch}{1.25}
\setlength{\tabcolsep}{1mm}{
\begin{tabular}{cccccc|cc}
\bottomrule[0.17em]
 & \textbf{\bm{$\mathcal{S}$}} & \textbf{Evolution} & \textbf{Phy. Con.} & \textbf{Global} & \textbf{Local} & \textbf{PSNR$\uparrow$} & \textbf{LPIPS$\downarrow$} \\ \hline\hline
1 & \textit{w./o. $S_{p}$} & \ding{55} & \checkmark & SA & ADAM &21.45  &0.2709  \\
2 & \textit{w./o. $S_{p}$} & \checkmark & \ding{55} & SA & ADAM &24.27  &0.1742  \\
3 & \textit{w./o. $S_{p}$} & \checkmark & \checkmark & SA & ADAM &25.50  &0.1598  \\
4 & \textit{w./o. $S_{p}$} & \checkmark & \checkmark & \textit{w./o.} & ADAM &23.59  &0.2015  \\
5 & \textit{w./o. $S_{p}$} & \checkmark & \checkmark & PSO & ADAM &24.71  &0.1675  \\
6 & \textit{w./o. $S_{p}$} & \checkmark & \checkmark & SA & DLS &23.76  &0.1655  \\ \hline
7 & \textit{w./o. $S_{aper}$} & \checkmark & \checkmark & SA & ADAM &25.00  &0.1307  \\
8 & \textit{w./o. $S_{FoV}$} & \checkmark & \checkmark & SA & ADAM &25.42  &0.1267  \\
9 & \textit{w./o. $S_{F}$} & \checkmark & \checkmark & SA & ADAM &25.11  &0.1682  \\
10& \textbf{\textit{all}} & \checkmark & \checkmark & \textbf{SA} & \textbf{ADAM} &\textbf{25.86}  &\textbf{0.1214}  \\ \hline
\end{tabular}

}
}
    \end{center}
\end{table}

\begin{table}[h!]
    \begin{center}
        \caption{\textbf{Ablations on DA Training.} ``$\dag$'' means that $\mathcal{L}_{DCP}$ is used on $\mathcal{D}_{S}$. $N_{T}$: Number of applied target images in DA training.}
        \label{tab:ab_da}
        \resizebox{0.5\textwidth}{!}
{

\renewcommand{\arraystretch}{1.25}
\setlength{\tabcolsep}{3mm}{
\begin{tabular}{cccc|cc}
\bottomrule[0.17em]
 & \bm{$N_T$} & \bm{$\mathcal{L}_{CAC}$} & \bm{$\mathcal{L}_{DCP}$} & \textbf{PSNR$\uparrow$} & \textbf{LPIPS$\downarrow$} \\ \hline\hline
zero-shot & 0 & \ding{55} & \textit{w./o.} &25.13  &0.1409  \\
1 & 0 & \checkmark & \textit{w./o.} &25.30  &0.1350  \\
2 & 0 & \checkmark & \textit{w.s.} 15$^\dag$ &25.19  &0.1337  \\
3 & 5 (1\%) & \checkmark & \textit{w.s.} 15 &25.35  &0.1262  \\
4 & \textbf{40 (5\%)} & \textbf{\checkmark} & \textbf{\textit{w.s.} 15} &\textbf{25.61}  &0.1260  \\
5 & 800 (100\%) & \checkmark & \textit{w.s.} 15 &25.49  &\textbf{0.1253}  \\
6 & 40 (5\%) & \checkmark & \textit{w.s.} 35 &25.51  &0.1257  \\
7 & 40 (5\%) & \checkmark & \textit{w.s.} 95 &25.49  &0.1264  \\ \hline
\end{tabular}

}
}
    \end{center}
\end{table}

\noindent\textbf{Ablations on Sampling Strategy.}
This ablation investigates how to sample proper lenses to construct an effective AODLib.
Tab.~\ref{tab:ab_sam} shows the zero-shot results of the pre-trained model under AODLibs with different sampling settings to intuitively reflect the quality of sampled LensLibs. 
It can be observed that a uniform $RMS$ distribution can eliminate the possible data preference, and more lens samples contribute to a larger coverage of LensLib.

\noindent\textbf{Ablations on EAOD.}
We explore the effectiveness of the components in EAOD as shown in Tab.~\ref{tab:ab_aod}.
Fixing $S_p$, experiments 1-4 illustrate the significance of the evolution mechanism, physical constraints, and global optimization for guaranteeing AODLib's wide and realistic coverage to develop a robust universal model.
Then, in terms of the implementation of the hybrid optimization strategy, we suggest the application of SA for global optimization and ADAM for local optimization to achieve better results (experiments 3, 5, 6).
\JQ{Beyond performance, compared with the other LensLibs in Table~\ref{tab:lib}, any combination of global and local optimization algorithms can produce a relatively high‑quality LensLib, indicating the advantage of the evolution mechanism in LensLib generation. Moreover, the local optimization algorithm, which determines the output multiple local optima, reveals a greater overall impact.}
In addition, experiments 3 and 7-10 reveal that each specification contributes to building a superior AODLib, where the absence of any one will limit the diversity of lens samples, resulting in a decline in model performance.

\noindent\textbf{Ablations on DA Training.}
Tab.~\ref{tab:ab_da} reports the ablations on the usage of target images and DCP loss during DA training.
From experiments 1-2, 4 and the zero-shot results, $\mathcal{L}_{DCP}$ can improve the CAC results when employed only on $D_S$, but incorporating target images with $\mathcal{L}_{DCP}$ delivers superior results with improvements of $0.42dB$ in PSNR and ${-}0.0077$ in LPIPS.
Furthermore, a few real-world captured target images ($1\%{\sim}5\%$) can enable effective DA training as shown in experiments 3-5, illustrating the flexibility of the OmniLens framework which can be easily implemented in real-world cases.
Additionally, experiments 4, 6, and 7 show that the window size (\textit{w.s.}) for calculating $\mathcal{L}_{DCP}$ reveals a minimal impact on the DA results, which is suggested to be set to $15$ for ensuring computational efficiency.

\section{Conclusion and Discussion}
\label{sec:conclusion}
\subsection{Conclusion} 
This paper introduces a flexible framework OmniLens for universal CAC based on pre-training on LensLib and LensLib-to-specific domain adaptation.
A pre-training model with strong generalization ability is first developed based on a lens database AODLib.
This is achieved by the proposed EAOD method for automatically designing large amount lens samples covering a diverse range of real-world aberration behaviors. 
The pre-trained model also reveals the potential to serve as a powerful pre-trained foundation.
Moreover, we develop an efficient domain adaptation method to deal with the specific lens with unknown lens descriptions based on the DCP property of optical degradation.
For the first time, DA is verified to be an effective solution to universal CAC, improving the pre-trained model only with fast fine-tuning (about $1$ hour) on a few easily accessible real-world captured images ($25$ images).

\subsection{\JQ{Discussion and Future Work}}
\PAR{\JQ{Enriching the LensLib.}}
Considering the complexity of optimization, the EAOD currently only accommodates spherical lens structures, leaving room for improvements in the coverage of AODLib. 
Introducing additional degrees of freedom in optimization, such as aspheric and diffractive elements, could yield design outcomes with richer aberration behaviors, which can hardly be achieved by spherical lenses. 
\JQ{Additionally, simulating manufacturing tolerances and assembly errors on optimized lens samples is highly valuable. These factors introduce stochastic, unknown aberration behaviors, making them a natural application case for blind aberration correction. Future work can center around incorporating more design specifications and perturbing LensLib samples within tolerance bands to diversify aberration distributions, thereby broadening dataset coverage and strengthening generalization for the pre-trained model.}

\PAR{\JQ{Designs for Model Paradigm.}}
The rich aberration behaviors of AODLib have not been fully leveraged during the training of our pre-trained model, as the model is merely fed with data from lenses with various aberrations in a random manner. We believe that appropriately classifying and annotating the aberration behaviors in AODLib, along with incorporating suitable prompts during model training, could enhance the generalization capability of the pre-trained models. In this context, we are particularly interested in utilizing prompt learning techniques in recent works of all-in-one image restoration.

\PAR{\JQ{Compensation for DCP Regularization.}} 
\JQ{While DCP is promising as an unsupervised constraint for CAC, training instability and darker outputs indicate room for improvement. A practical path is to augment DCP with auxiliary low-level vision priors, as commonly done in image dehazing~\cite{chen2021psd,li2019semi}.
For example, the bright channel prior can compensate for DCP failures in sky and specular regions, and the gradient sparsity prior, which is often expressed as Total Variation Loss (TV Loss), can further suppress noise and artifacts brought by the unsupervised DCP constraint.}

\bibliographystyle{SELF-cas-model2-names}
\bibliography{ref}

\section*{Funding}
Zhejiang Provincial Natural Science Foundation of China (Grant No. LZ24F050003). 
Yuyao City Enterprise Innovation Consortium Project (2024YYS220008). 
National Natural Science Foundation of China (NSFC) under Grant No. 62473139.
Hunan Provincial Research and Development Project (Grant No. 2025QK3019).
Open Research Project of the State Key Laboratory of Industrial Control Technology, China (Grant No. ICT2025B20).
State Key Laboratory of Autonomous Intelligent Unmanned Systems (the opening project number ZZKF2025-2-10).

\end{document}